\documentclass[prb,twocolumn,showpacs,floatfix,,amsmath,amssymb,superscriptaddress]{revtex4}
\usepackage{amsfonts}
\usepackage{stmaryrd}
\usepackage{bbm}
\usepackage{mathrsfs}
\usepackage{tipa}
\usepackage{amssymb}
\usepackage{txfonts}
\usepackage{graphicx}
\usepackage{dcolumn}
\usepackage{epstopdf}
\usepackage[colorlinks,linkcolor=blue,urlcolor=blue,citecolor=blue]{hyperref}
\usepackage{multirow}
\usepackage{subfigure}

\begin{document}
\newcommand*{\cm}{cm$^{-1}$\,}
\newcommand*{\Tc}{T$_c$\,}

\title{Optical study of phase transitions in single-crystalline RuP}
\author{R. Y. Chen}
\affiliation{International Center for Quantum Materials, School of Physics, Peking University, Beijing 100871, China}
\author{Y. G. Shi}
\affiliation{Beijing National Laboratory for Condensed Matter
Physics, Institute of Physics, Chinese Academy of Sciences,
Beijing 100190, China}
\author{P. Zheng}
\affiliation{Beijing National Laboratory for Condensed Matter
Physics, Institute of Physics, Chinese Academy of Sciences,
Beijing 100190, China}
\author{L. Wang}
\affiliation{Beijing National Laboratory for Condensed Matter
Physics, Institute of Physics, Chinese Academy of Sciences,
Beijing 100190, China}
\author{T. Dong}
\affiliation{International Center for Quantum Materials, School of Physics, Peking University, Beijing 100871, China}
\author{N. L. Wang}
\affiliation{International Center for Quantum Materials, School of Physics, Peking University, Beijing 100871, China}
\affiliation{Collaborative Innovation Center of Quantum Matter, Beijing, China}

\begin{abstract}
RuP single crystals of MnP-type orthorhombic structure were synthesized by the Sn flux method. Temperature-dependent x-ray diffraction measurements reveal that the compound experiences two structural phase transitions, which are further confirmed by enormous anomalies shown in temperature-dependent resistivity and magnetic susceptibility. Particularly, the resistivity drops monotonically upon temperature cooling below the second transition, indicating that the material shows metallic behavior, in sharp contrast with the insulating ground state of polycrystalline samples. Optical conductivity measurements were also performed in order to unravel the mechanism of these two transitions. The measurement revealed a sudden reconstruction of band structure over a broad energy scale and a significant removal of conducting carriers below the first phase transition, while a charge-density-wave-like energy gap opens below the second phase transition.

\end{abstract}


\maketitle

\section{introduction}
The superconductivity discovered in LaO$_{1-x}$F$_x$FeAs\cite{doi:10.1021/ja800073m} in 2008 has triggered tremendous enthusiasm in the scientific community to study the unconventional iron-based superconductors. Subsequent investigations reveal that both the magnetic order in parents and superconductivity in doped compounds originate from the FeAs layers, and the relative distance and angles between Fe and As atoms are crucial to the superconducting transition temperatures\cite{Lee2008,Kuroki,Mizuguchi}. Therefore, it is natural to inspect the properties of iron monoarsenide (FeAs)-type materials, which are supposed to provide useful information for the paring mechanism in iron-based superconductors, concerning its relatively simple constitution. However, even though the FeAs single crystal has been known for decades, no superconducting evidences are hitherto discovered in this compound \cite{hagg1928,rundq,RuAs1961,Mossbauer,Lyman,CrFeAs}. Distinct from the so-called ``11'' type superconductor FeSe \cite{FeSeHsu,FeSe2008} or the FeAs layers in other iron-based superconductors, FeAs compound crystalizes in a MnP-type orthorhombic structure \cite{selte1969} and shows double-helical magnetic structure at low temperatures \cite{jones1967,yazuri}, which makes it irrelevant to iron-based superconductors.
Nevertheless, its isostructural cousin CrAs was reported to exhibit superconductivity under external high pressure very recently \cite{CrAs2014}, which is of significant importance. Not only does this groundbreaking work report the first Cr-based superconductor, it also draws much attention to the transition metal monopnictides once again, most members of which possess the same orthorhombic structure with the \textit{pnma} space group, such as MnAs, RuAs, CrP, MnP, FeP, CoP, and RuP etc. \cite{rundq,MnAs1964,CrP1981,selte1972}.

Notably, a recent research \cite{PhysRevB.85.140509} on RuP polycrystals claimed that this compound experiences two successive structural phase transitions at 330 K ($T_1$) and 270 K ($T_2$), respectively. Moreover, the lower-temperature transition is considered as a metal to nonmagnetic insulator transition. Of most importance, superconductivity could be induced by Rh doping, which gradually suppresses both of the two phase transitions. Together with CrAs, this discovery raised the possibility of the MnP-type materials being promising candidates for exploring novel superconductors, even though they have totally different ground states. CrAs undergoes a first-order antiferromagnetic phase transition near 270 K, which could be suppressed and gives way to superconductivity, whereas RuP turns to insulator below $T_2$. Shortly after, photoemission spectroscopy\cite{RuPXPS} was employed to resolving the mechanism of the two exotic phase transitions, suggesting that the Ru valence is +3 with $t^5_{2g}$ configuration, and the two consecutive transitions would stem from Peierls-like transitions in $xy$ and $yz/zx$ orbital channels, respectively.

We have successfully grown large-size (5 $ \times$ 1 $\times$ 1 mm) single-crystalline RuP. Two distinct phase transitions have been confirmed by x-ray diffraction and transport measurements, occurring around $T_1$ and $T_2$, respectively. A pronounced hysteresis is observed in both temperature-dependent resistivity and magnetic susceptibility measurements for the first phase transition, indicative of first-order type. Below $T_2$ the compound shows metallic response, which is in stark contrast with the previously reported metal-insulator transition \cite{PhysRevB.85.140509}. To further address this issue, we performed infrared spectroscopy on the compound. The measurement revealed a dramatic reconstruction of the band structure over a broad energy scale and a significant reduction of conducting carriers below the first phase transition, while the formation of a partial charge density wave (CDW)-like energy gap is observed for the second phase transition.

\section{experiment and results}
\begin{figure}[htbp]
\includegraphics[width=7cm]{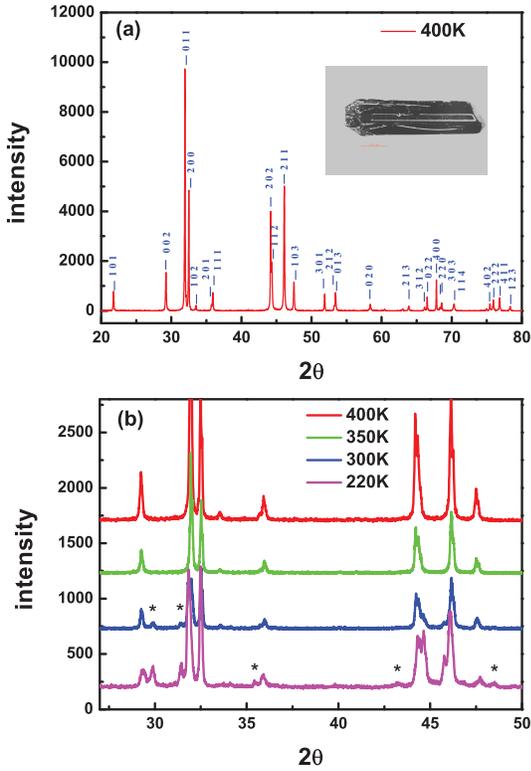}
\caption{(Color online) (a) X-ray diffraction patterns of RuP compound at 400 K, and (b) temperature-dependent x-ray diffraction patterns. The peaks denoted by star marks are the emerging ones compared to higher temperatures.}
\label{Fig:XRD}
\end{figure}

Single crystals of RuP were synthesized by the Sn flux method. The starting composition of RuP$_{1.02}$Sn$_{25}$ was placed in a crucible and sealed in an evacuated quartz tube, which was afterwards heated to 1150 $^{\circ}$C, and then cooled to 950 $^{\circ}$C at a rate of 2 $^{\circ}$C/h. The 2\% excess of phosphorous was added so as to compensate for the loss due to volatilization. Prismatic single crystals of RuP [as shown in the inset of Fig.\ref{Fig:XRD} (a)] were obtained after eliminating Sn flux by soaking in concentrated hydrochloric acid for hours.

Several pieces of the obtained single crystals were ground up in order to conduct powder x-ray diffraction (XRD) experiments. Figure \ref{Fig:XRD}(a) shows the XRD patterns taken at 400 K, which can be well refined by the reported MnP-type orthorhombic structure. The fitted lattice parameters are \textit{a} = 5.54 \AA, \textit{b} = 3.17 \AA, and \textit{c} = 6.14 \AA, very close to the previous report \cite{rundq}. There are almost no impurity peaks within the accuracy of measurement, which assures the high quality of our samples. Temperature-dependent XRD measurements were also performed as shown in Fig. \ref{Fig:XRD} (b). The patterns at 350 K have no perceptible discrepancy with those at 400 K. As temperature decreases, however, several additional peaks become observable at 300 K, indicative of a structural phase transition. In order to reveal the underlying mechanism of this transition, it would be of great importance to identify the exact lattice structure of the compound at room temperature, which is unfortunately unavailable yet. As for 220 K, more extra peaks develop at this temperature, which evidences another structural transition. Besides, there are also weak shifts of positions and relative intensities among the peaks, indicating the complicated variation of lattice structure.
\begin{figure}[htbp]
\includegraphics[width=5.5cm]{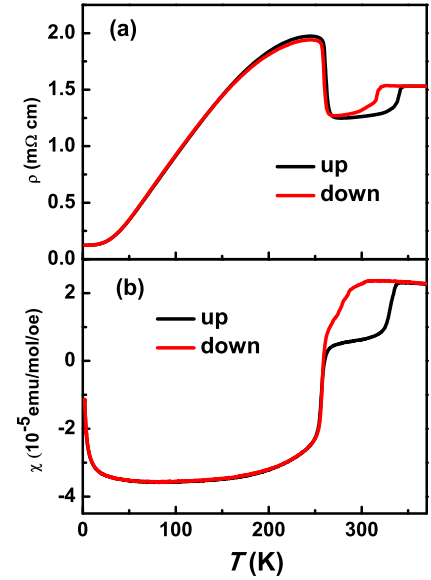}
\caption{(Color online) Temperature dependent (a) resistivity and (b) magnetic susceptibility. The red and black lines indicate the cooling and warming process, respectively.}
\label{Fig:res}
\end{figure}
The temperature-dependent resistivity $\rho (T)$ was measured using the standard four-probe method in a physical property measurement system (PPMS) with current along the \textit{a} axis, and the results are presented in Fig. \ref{Fig:res} (a). Above 350 K, the resistivity is almost temperature independent, whereas below 350 K it shows two individual anomalies. As temperature decreases, $\rho (T)$ drops sharply around 320 K, corresponding with the high-temperature structural transition. Then it decreases in a milder way until 270 K, where a sudden upturn appears, signaling the second phase transition. Huge hysteresis in the cooling and heating cycles was observed for the first anomaly, implying that the transition is of first order. Remarkably, $\rho (T)$ shows metallic behavior at lower temperatures until 2 K, deviating violently from the previous report \cite{PhysRevB.85.140509}, in which a metal-insulator transition is claimed to occur at $T_2$. This huge discrepancy might arise from the differences between qualities of samples. The insulating properties were obtained by polycrystalline samples, which is very likely to be affected by a grain boundary effect or other defects. On the contrary, since our data were collected on highly pure single crystals, they represent the intrinsic properties of RuP.

Magnetic susceptibility $\chi(T)$ was measured in a superconducting quantum interference device (SQUID) vibrating-sample magnetometer (VSM) with a 1-T magnetic field applied parallel to the \textit{a} axis, as displayed in Fig. \ref{Fig:res} (b). It also exhibits two obvious anomalies, one of which shows enormous hysteresis, agreeing well with the resistivity measurement. Above the first transition temperature $\chi(T)$ is positive with a quite small absolute value, implying a Pauli paramagnetic susceptibility. Upon temperature cooling, $\chi(T)$ reduces monotonically and drops almost discontinuously from positive to negative at $T_2$, which is reminiscent of Larmor diamagnetic susceptibility. Pauli magnetism is generally in proportion to the density of states of free carriers, while Larmor diamagnetism is linked with the orbital motion of electrons in the closed shells. On this account, the transformation from Pauli paramagnetic to Larmor diamagnetic susceptibility reflects sudden reductions of density of states of free carriers across both transitions. It is noted that the susceptibility is almost temperature independent below the second phase transition, except for the small upturn at very low temperature, which could be attributed to the imperfection of the crystal. This result demonstrates a nonmagnetic ground state of RuP compound.
\begin{figure}[t]
\includegraphics[width=7cm]{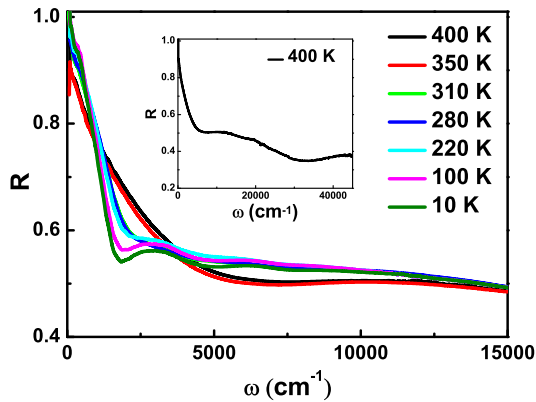}
\caption{(Color online) Temperature-dependent optical reflectivity $R(\omega)$ below 18 000 \cm . The inset displays $R(\omega)$ in a large energy scale range of up to 45 000 \cm at 400 K.}
\label{Fig:ref}
\end{figure}

The optical reflectance of the as-grown plane, which was determined to be the (0 1 1) plane, was measured on a combination of Bruker Vertex 80-V and 113-V Fourier transform spectrometers in the frequency range 40--25 000 \cm. Grating-type spectrometry was also employed to probe higher energy responses up to 45 000 \cm. An \textit{in situ} gold and aluminum overcoating technique was used to get the reflectance $R(\omega)$. The real part of optical conductivity $\sigma_1(\omega)$ is obtained by the Kramers-Kronig transformation of $R(\omega)$, which was extrapolated by a Hagen-Rubens relation in the low-energy side, then by $\omega^{-0.2}$ in the high-energy side until 100 000 \cm, and by $\omega^{-4}$ for higher energies.

The inset of Fig. \ref{Fig:ref} displays the reflectance spectrum over the whole measurement energy scale from 40 to 45 000 \cm at 400 K, where $R(\omega)$ drops monotonically from unit at zero frequency and reaches a minimum at around 5000 \cm, usually referred to as a ``screened" plasma edge. The main panel shows the spectra at various temperatures in an expanded scale from 0 to 18 000 \cm. It is seen clearly that $R(\omega)$ approaches unit at zero frequency and increases as temperature decreases at very low frequency, revealing the metallic behavior of the RuP single crystal. As temperature decreases, the spectra change dramatically across the first transition. The spectra above and below the first transition are separated into two groups. For compounds experiencing the first-order phase transitions, e.g., IrTe$_2$ \cite{Fang2013} and BaNi$_2$As$_2$ \cite{BaNIAs}, we
commonly observe sudden spectral change over broad energy scales across the phase transitions, arising from the reconstruction of band structures. At the same time, the plasma edge gets much sharper and shifts to lower frequency. Provided the effective mass of free carriers remains unchanged, the red-shift of the plasma edge frequency, which is in proportional to $(n/m^*)^{1/2}$, demonstrates a sudden reduction of the carrier density. Through the second phase transition, $R(\omega)$ experiences only quite delicate variations, that is, a minor spectral weight transformation from low energy below 2000 \cm to higher energy. In addition, a weak hump develops at around 3200 \cm, which grows more pronounced upon temperature cooling.

\begin{figure}[t]
\includegraphics[width=7cm]{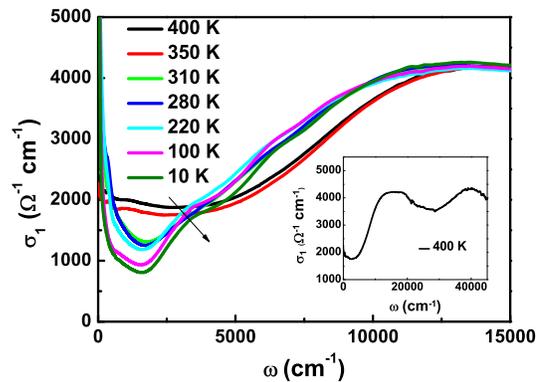}
\caption{(Color online) Temperature-dependent optical conductivity below 15 000 \cm. The inset shows $\sigma_1(\omega)$ up to 45 000 \cm at 400 K. }
\label{Fig:con}
\end{figure}
The evolution of physical properties across the two phase transitions could be more distinctively resolved by the optical conductivity $\sigma_1(\omega)$. Figure 4 shows $\sigma_1(\omega)$ below 15 000 \cm for different temperatures in its main panel, while the inset present $\sigma_1(\omega)$ in a large energy scale ranging up to  45 000 \cm at 400 K. The Drude-like component is rather broad above $T_1$, indicating that the charge carriers experience rather strong scattering. Across the first phase transition, the Drude feature sharply narrows and shifts to lower frequencies. It can be found that the spectral weight below 3500 \cm is severely suppressed, part of which moves to lower frequencies while the rest to much higher energies up to 12000 \cm (1.5 eV). The conductivity spectra are also separated into two groups above and below $T_1$, indicating the band structure reconstruction. The result yields further evidence for the first-order phase transition at $T_1$, being consistent with the resistivity and magnetic susceptibility measurements. Below $T_2$, the Drude component becomes further narrowed. Notably,
a broad peak emerges at around 3200 \cm, the central frequency of which moves slightly to higher energies with decreasing temperature, as indicated by the arrow in Fig. \ref{Fig:con}. The result suggests the formation of a partial energy gap in the density of states. It is thus expected that the second phase transition at $T_2$ corresponds to a development of broken symmetry state, like charge order or charge density wave.

\begin{figure*}
  \centering
  \includegraphics[width=15cm]{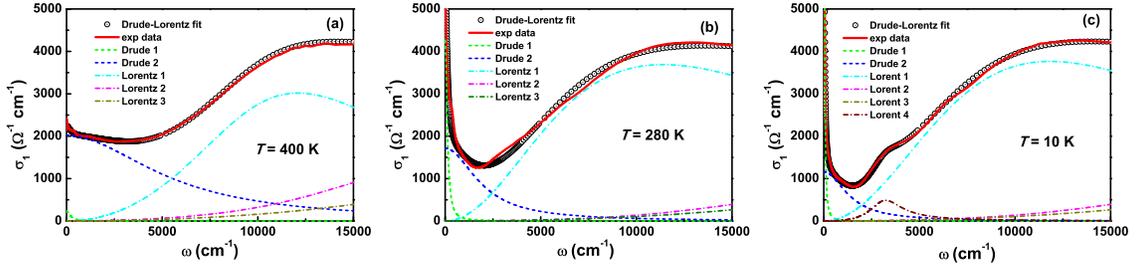}\\
  \caption{(Color online) The Drude-Lorentz fitting of optical conductivity $\sigma_1(\omega)$ for (a) \textit{T} = 400 K, (b) \textit{T} = 280 K, and (c) \textit{T} = 10 K.  }
  \label{Fig:fit}
\end{figure*}
\section{discussion}

In order to get more information and interpret the data quantitatively, we use the Drude-Lorentz model to decompose the optical conductivity $\sigma_{1}(\omega)$:
\begin{equation*}
\epsilon(\omega)= \epsilon_{\infty}-\sum_{s}{\frac{\omega_{ps}^2}{\omega^2+i\omega/\tau_{Ds}}}+ \sum_{j}{\frac{S_j^2 }{\omega_j^2-\omega^2-i\omega/\tau_j}}. 
\end{equation*}
Here, $\varepsilon_{\infty}$ is the dielectric constant at high energy; the middle and last terms are the Drude and Lorentz components, respectively. We found that the conductivity above $T_2$ can be well reproduced
by two Drude and three Lorentz terms as shown in Figs. \ref{Fig:fit}(a) and \ref{Fig:fit}(b). The two Drude terms represent free carriers from different Fermi surfaces, whereas the Lorentz components are used to describe the excitations across energy gaps and interband transitions. The second Drude term has a much larger spectral weight. It is also much broader than the first one, implying a
\begin{figure}[b]
  \centering
  \includegraphics[width=6.5cm]{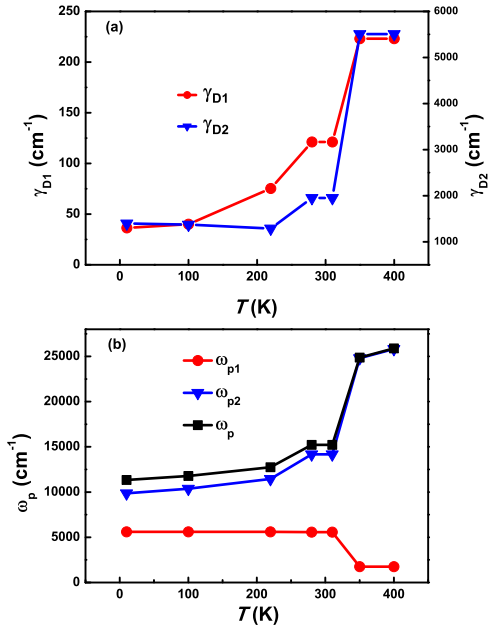}\\
  \caption{(Color onlin) The temperature-dependent (a) scattering rate $\gamma_{D1}$, $\gamma_{D2}$ and (b) plasma frequency $\omega_{p1}$, $\omega_{p2}$, $\omega_{p}$. }
  \label{Fig:parameter}
\end{figure}
much larger scattering rate $\gamma_D$ ($ 1/\tau_D$). As temperature decreases, the Drude terms shrink substantially and an additional Lorentz term (Lorentz 4) could be resolved at around 3200 \cm below $T_2$. Apart from that, the rest of the Lorentz terms are highly stable in spite of temperature change, which accordingly originate from high-energy interband transitions.

Through the first structural transition, both of the scattering rates $\gamma_{D1}$, $\gamma_{D2}$ drop extremely sharply, as depicted in Fig. \ref{Fig:parameter} (a). Simultaneously, the plasma frequency $\omega_{p1}$ and $\omega_{p2}$ also experience a huge shift, but evolve in an opposite manner, as shown in Fig. \ref{Fig:parameter} (b). As is well known, the square of the plasma frequency is proportional to $n/m^*$, where $n$ and $m^*$ are the number of free carriers and effective mass of electrons, separatively. Therefore, we can conclude that majority of the strongly scattered free carriers are lost during this transition while the number of those with smaller $\gamma$ gains a mount of population due to the band structure reconstruction. To get an overall perspective, we use $\omega_p=(\omega^2_{p1}+\omega^2_{p2})^{1/2}$ to represents the general plasma frequency, which decreases from 24 900 \cm to 15 200 \cm across the first transition. The analysis indicates that a majority of itinerant charge carriers are lost in this process.

The variation of fitting parameters through the second phase transition seems to be much more delicate. Both the scattering rate and plasma frequency continue to drop but in a gentler way. The reduction of $\omega_{p}$ leads to the further decreasing of magnetic susceptibility $\chi(T)$, which transforms from Pauli paramagnetism to Larmor diamagnetism. It is worth noting that the magnitude of the decreasing of $\omega_p$ across the first transition is much larger than the second one; however, the dropping of $\chi(T)$ is quantitatively comparable. This contradiction is probably ascribed to the change of effective mass $m^*$, which contributes to the plasma frequency as well. Moreover, the sudden increase of resistivity at $T_2$ could also stem from the decrease of free carriers. Below $T_2$, all the parameters are almost constant except for the scattering rate $\gamma_{D1}$. It decreases as temperature decreases, in accord with a good metal behavior.

The emergence of the new peak feature at the mid-infrared region (about 3200 \cm or 0.4 eV) below the second phase transition is likely an indication of formation of a CDW energy gap. Its central frequency evolves towards a higher energy level, which is qualitatively in accordance with the BCS theory. In an earlier study, Hirai \textit{et al}. \cite{PhysRevB.85.140509} excluded the existence of CDW based on the finding of an insulating ground state, since in such a three-dimensional system a CDW order is unexpected to cause an insulating ground state. However, it is clearly elaborated here that the ground state of RuP single crystal is actually metallic, rationalizing the appearance of a partially gapped CDW order. The gap energy is identified by the center frequency of the corresponding Lorentz peak in $\sigma_1(\omega)$ to be $2\Delta$ = 400 meV, which yields $2\Delta/k_BT$ = 16.8. Although the value is much larger than the weak-coupling theory prediction, it is still in the range among many strongly coupling CDW systems \cite{affang}, which could be explained by the microscopic strong-coupling theory developed by Varma and Simons \cite{PhysRevLett.51.138}.
It is noted that a recent ultraviolet photoemission spectroscopy study indicated that the surface of RuP polycrystalline samples remains metallic through the second transition. This experiment seems to support our result that the intrinsic ground state is actually metallic. Based on our results, we could infer that the first structural phase transition is of the nature of a first-order phase transition, which leads to dramatic reconstruction of the electronic band structure over a large energy scale, while the second structural phase transition is likely due to the formation of CDW order where the structural change is not dramatic but only small change in some bond lengths.
Apparently, further studies are still needed to resolve precisely the structural changes across both phase transitions.

\section{summary}

In conclusion, our measurements on single-crystalline RuP elaborate that the compound undergoes two structural phase transitions. The higher-temperature one shows enormous hysteresis and thus is of first-order type. The optical study further revealed the band structural reconstruction over a broad energy scale. A majority of the free charge carriers are removed due to this band structure reconstruction, yet the concurrent reduction of scattering rate yields a better metallic behavior. On the other hand, the second structural phase transition is linked with the formation of a partial energy gap. It is likely to be driven by a CDW instability. The compound remains metallic below the second structural phase transition, which is in sharp contrast to the report in an earlier study based on polycrystalline samples. Since the MnP-type materials are very likely to be a new superconducting family, our research on RuP single crystals has extended this realm and provided an important foundation for further investigations.

\begin{center}
\small{\textbf{ACKNOWLEDGMENTS}}
\end{center}

This work was supported by the National Science Foundation of
China (Contracts No. 11120101003, and No. 11327806), and the 973 project of the Ministry of Science and Technology of China (Contracts No. 2011CB921701, No. 2012CB821403).

\bibliographystyle{apsrev4-1}
  \bibliography{RuP}

\end{document}